\documentclass[12pt,a4paper]{article}
\usepackage{amsmath}
\usepackage{epsfig}
\usepackage{amstext}
\usepackage{graphicx}
\usepackage{theorem}
\usepackage{tabularx}
\usepackage{enumerate}
\usepackage{amssymb}
\usepackage{amscd}

\newtheorem{theorem}{Theorem}

\newtheorem{lemma}[theorem]{Lemma}

\newtheorem{remark}[theorem]{Remark}

\begin{document}

\title{On multisolitonic decay behavior of internal gravity waves}
\author{A. A. Halim $^{1}$, S.P.Kshevetskii $^{2}$, S.B. Leble$^{3}$ \\
%EndAName
$^{1,3}$ Technical University of Gdansk, \\
ul. G. Narutowicza 11/12, 80-952 Gdansk, Poland. \\
$^{2}$ Kaliningrad state University, Kaliningrad, Russia\\
$^{3}$leble@mif.pg.gda.pl}
\maketitle

\begin{abstract}
We claim that changes of scales and fine-structure could increase
from multisoliton behavior of internal waves dynamics and,
further, in the so-called "wave mixing". We consider
initial-boundary problems for Euler equations with a stratified
background state that is valid for internal water waves. The
solution of the problem we search in the waveguide mode
representation for a current function. The orthogonal
eigenfunctions describe a vertical shape of the internal wave
modes and satisfy a Sturm-Liouville problem for the vertical
variable. The Cauchy problem is solved for initial conditions with
realistic geometry and magnitude. We choose the geometry and
dimensions of the McEwan experiment with the stratification of
constant buoyancy frequency. The horizontal profile is defined by
numerical solutions of a coupled Korteweg-de Vries system. The
numerical scheme is proved to be convergent, stable and tested by
means of explicit solutions for integrable case of the system.
Together with the solution of the Sturm-Liouville problem it gives
the complete internal waves profile.
\end{abstract}

\section{Introduction}

There are papers (see S. B. Leble (1991) for a review) in which
the problem of nonlinear waves evolution in stratified medium is
in some way connected with soliton theory, see, e.g., Novikov et
al (1984). We mean the conditions in which the physical (or
environmental) conditions need account of nonlinearity and
dispersion which joint action is of essential (sometimes -
inevitable) importance. Evolution of internal gravity waves (IGW)
in geophysical hydrodynamics and its further decay to turbulent
spots is important example of such phenomena, Pedlosky (1979). The
scenarios of this transition intensely discussed in literature of
atmosphere and ocean physics Gavrilov, Fukao, Nakamura, Tsuda
(1996). There are also publications about laboratory experiments
devoted to IGW dynamics in tanks Segur, Finkel (1985). The problem
have a link to a general problems of hydrodynamics as illustrated
by Zeitunyan (1999). It is the Euler equations used as a basis of
a description in which modes with different scales are generated
within the process of a wave propagation over an initial density
stratification taking into account its possible dynamic changes.
This question relates to the behavior of a solution within the
limiting procedure when a dispersion coefficient go to zero. In
our case of internal waves it means the zero buoyancy frequency
limit ($N \rightarrow 0 $ ) that in turn is proportional to the
vertical density gradient. Let us mention also investigations of
similar problems with a small dispersion parameter initiated by
Lax and Levermore (1983) for the
 model of a single Korteweg-de Vries (KdV) equation. The
 dispersion in such model is described by the term with third
 derivative by a coordinate (x).

In this paper we study an evolution of a disturbance from an
initial condition that model an excitation of the internal wave in
the geometry and scales of McEwan (1983 a,b) experiments. We
believe that the fine structures that appear in density
distribution  may be explained by a model based on a system of
coupled Koreweg-de Vries  (cKdV) equations. This system arises for
coefficient functions related to a functional basis of transversal
coordinate. It demonstrates the behaviors of solutions that have
many features of multi-solitonic decay (see, again, Novikov et al
(1984)) The wavetrains contain chains of such fluctuations of
basic wave variables that propagate with velocity up to the most
quick one. Our investigations show: interaction between the basic
transverse guide modes does not destroy the general picture; only
some phase changes appear. Of course the system of cKdV has higher
order in time derivatives, the opposite directed wave train is
launched, but the wave also has the multisoliton form.

The derivation of the system of cKdV may be considered as the
direct application of Galerkin - Bubnov projection procedures that
is successfully used in numerical simulation of initial-boundary
problems with a good mathematical justification (the basis of this
method are in Gottlieb and Orszag (1989), see, also Moser (1966)).
After the variables division in the linearized problem some
spectral problem appears. Its eigen functions of transverse
variables form a basis. Hence the (infinite) 1+1 cKdV system is
equivalent to the 2+1 basic Euler system under consideration
within some class of initial conditions. While numeric simulation
delivering we cut this infinite system restricting ourselves by
some mode number, comparing few such examples. We, however believe
that further extending of the system does not change the general
behavior of the wavetrains. We plan checking of the hypothesis in
the forthcoming paper but more concentrate now on the cKdV model
that is also interesting by itself (e.g.  Hirota R., Satsuma J.
(1981) , A. Perelomova  (2000)). It was studied mathematically via
many approaches as described recently by Ayse Karasu (1997). There
are integrable examples (R. Dodd, A.Fordy (1982)): it helps to
test the numerical scheme to be  introduce in this article.

For such systems we construct a  finite-difference scheme (Sec. 3)
and use it for numerical integration. The theorems about
convergence and stability we formulate and prove here confirm the
statements we made. The computations description we place in the
section 4 and the proofs in Appendix.

In this article we do not touch the mixing problems in its full
extent as well as interaction with the stratification mode
(background or stationary state) itself. Such problems solution
need account of boundary conditions and simulations for a longer
time intervals. Account of such ingredients in the model could
shade the main phenomena we intend to demonstrate. Therefore we
concentrate attention of a reader on the initial stage of the wave
field development. Already within this initial stage period the
horizontal multi-solitonic fine structure prevails: it seems very
important. Our general approach and the numerical scheme power
allow to consider longer time to account reflections phenomena. A
comparison of the evaluated stream function fields with ones of
the McEwan experiments show many similar features, that, seems to
verify the model from the point of the real fluid dynamics.

\section{Formulation of the problem}

\subsection{Laboratory experiment of McEwan}
Experiment were made using a rectangular plate-glass tank with
dimensions shown in figure (1) below. This tank is filled by a
linearly stratified salt solution. A wave-forcing dismountable
paddle swung about horizontal lateral axis ia used \cite{ME2}.
\begin{center}
\epsfig{file=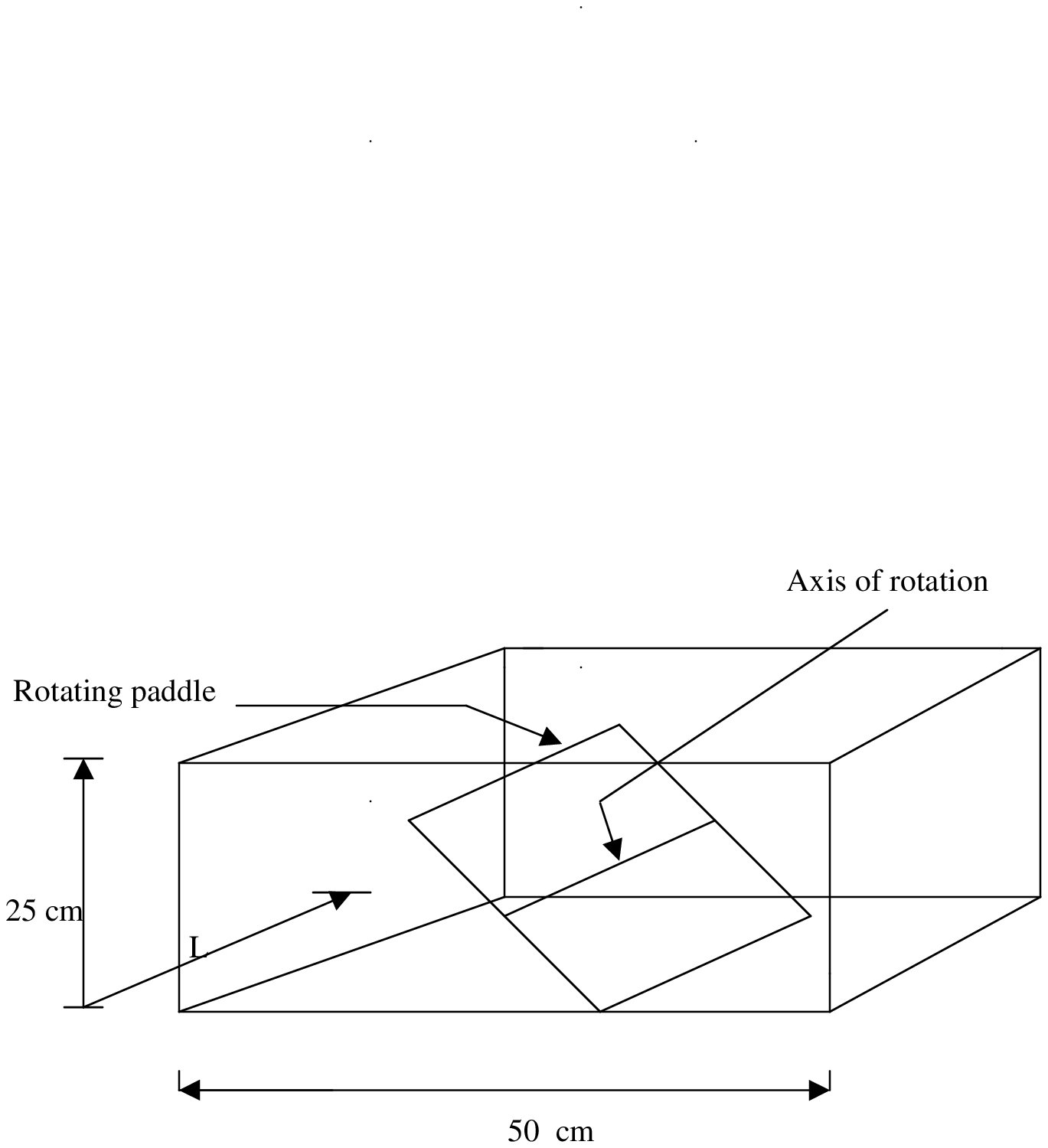, height=6.5cm, width=12  cm ,clip=,angle=0}\\
Tank filled with a continuous stratified ambient state water\\
Fig. (1) A. D. McEwan experimental configuration.
\end{center}
Bring the fluid to a state of incipient breaking and photographing
the evolution of these breaking events induced by internal wave
action. McEwan gives his
observation \cite{ME1}  along the following stages:\\
$(a)$ Overturning.\\
$(b)$ Development of interleaving microstructure.\\
$(c)$ Static stability is restored but microstructure is
preserved.\\
$(d)$ Gravitation to an equilibrium has changed the surrounding
density profile between extremum isopycnals.\\

We simulate only tile stage (b) above to show the obvious plots
for the mixing phenomena.

\section{Theoretical description and mathematical problem.}
Consider the basic system of Euler equations for internal water
waves in two dimensions (xz) with a stable stratified ambient
state and the buoyancy frequency $N(z)$.   \smallskip
\begin{equation} \label{Eul1}
\begin{array} {cc}
u_{x}+w_{z}=0 \\
\rho _{o}\,u_{t}=-\rho _{o}\left( \underline{v},\underline{\nabla
}\right) u-p_{x} \\
\rho _{o}\,w_{t}= \rho _{o}\left( \underline{v},\underline{\nabla
} \right) w-p_{z}-\rho ^{^{\prime }}\,g \\
T_{t}^{^{\prime }}+w \overline{T}\,_{z}\,=-\left( \underline{v},%
\underline{\nabla }\right) T^{^{\prime }}
\end{array}
\end{equation}
where u, v are velocity components, $\rho _{o}\,$\ is the density,
$p$ is the pressure, $\rho ^{^{\prime }}\,g$\thinspace\ is the
body force due to stratifications, $\overline{T}\,_{z}$ is the
vertical background temperature gradient and $T^{^{\prime }}$ is
the temperature variables.

In the commonly accepted approximations (incompressibility,
Boussinesq approximation, etc.) the solution of the system
(\ref{Eul1}) is constructed as the following representation for
the current function$\psi (x,z,t)$, (See, for details Appendix A)
%can be written in terms
 %approximation
\begin{equation}  \label{currentpsi}
\psi(z,x,t) =\sum\limits_{n=1}^{L} Z^{n} (z) \theta^{n}(x,t),
\end{equation}
where $Z^{n} (z)$ are solutions of the correspondent
Sturm-Liouville problem (Equation (\ref{separ}) in Appendix A)
\begin{equation}  \label{sturmliouv}
Z_{zz} +\frac{N^{2}}{c_n^{2}} Z=0, \, Z(0) = Z(h) = 0,
\end{equation}
and describe a vertical shape of the wave modes. The linear
propagation velocities, $c_n$, play the role of eigenvalues. The
series (\ref{currentpsi}) may be considered in the context of the
Galerkin - Bubnov method. (for a basis of this method see, for
example, Gottlieb and Orszag (1989)). The coefficient functions
$\theta^n(x,t)$ are solutions of the coupled KdV system (equation
(\ref{ckdv2}) in appendix A )
\begin{equation} \label{ckdv1}
\theta _{t}^{n}+c_{n}\theta _{x}^{n}+\sigma \underset{m,k}{\sum
\,g_{m,k}^{n}}\theta ^{m}\theta _{x}^{k}+\beta ^{2}d_{n}\,\theta
_{xxx}^{n}=0,
\end{equation}
  $\sigma ,\,\beta ^{2}\,$are scale parameters. Each solution $\theta ^{n}$ (x, t) stands for n-th mode wave
amplitude as a function of the space $x\in(-\infty,\infty)$ and
time t coordinates. Here the constants $g_{mkn},\,d_{nm}$ are
nonlinear and dispersion constants. The coefficients
$g_{mkn},\,d_{nm}$ have been  derived by Leble (1991) to have the
formulae (equation (\ref{coef2}) in appendix A )
\begin{equation} \label{coef1}
g_{m,k}^{n}=\frac{\sigma
N^{2}c_{n}^{2}}{2}\overset{h}{\underset{-h}{\int }}\left[ \left(
\frac{-1}{c_{m}^{2}}+\frac{2}{c_{k}^{2}}\right) Z^{k}Z_{z}^{m}-
\frac{1}{c_{m}c_{k}} Z^{m}Z_{z}^{k}\right]
z^{n}dz,\,d_{n}=\frac{c_{n}^{3}}{2N^{2}}
\end{equation}
We specify the problem by  zero boundary conditions for $\psi$
with respect to the variable $z$ as follows from
(\ref{sturmliouv}). Physically it corresponds to the zero value of
the vertical velocity components at the upper (z = h) and lower
boundaries  (z = 0) levels. We select a localized along x axis
initial condition described by a smooth enough function that model
the paddle motion as in the experiments of McEwan (1983a). The
function is also chosen antisymmetric along $x$ in relation to the
paddle axis centered in the middle of the tank. The time intervals
of simulations are taken such that the initial disturbance decays
essentially but does not reach the boundaries.

In the simplest model to be presented here we consider the case of constant
buoyancy frequency. In the concrete numerical calculations the choice of the
constants is the following $N=1.23\,\ s^{-1}$\,\ for a stratified salted
water in a rectangular tank 50 cm long filled to a depth 25 cm (Ref. McEwan
(1983a)).

\section{Numerical method}

The numerical scheme is based on the solution of the nonlinear finite
(difference) system

\begin{gather}
c\left( {\left( \theta ^{n}\right) _{i}^{j+\frac{1}{2}}-\left( \theta
^{n}\right) _{i}^{j}}\right) /\frac{\tau }{2}+c_{n}\left( {\left( \theta
^{n}\right) _{i+1}^{j}-\left( \theta ^{n}\right) _{i-1}^{j}}\right) /{2h}%
+\sum\limits_{k,m}g_{mkn}\left( \theta ^{m}\right) _{i}^{j}\left( {\left(
\theta ^{k}\right) _{i+1}^{j}-\left( \theta ^{k}\right) _{i-1}^{j}}\right) /{%
2h}  \label{ds} \\
+\left( d_{n}-{c_{n}h^{2}}/{6}\right) \left( {\left( \theta ^{n}\right)
_{i+2}^{j}-2\left( \theta ^{n}\right) _{i+1}^{j}+2\left( \theta ^{n}\right)
_{i-1}^{j}-\left( \theta ^{n}\right) _{i-2}^{j}}\right) /{\ 2h^{3}}=0,
\notag
\end{gather}
where $n$, $m$, $k$ are the modes numbers; $i$ and $j$ are discrete space
and time variables respectively. The time step is denoted by $\tau $ while
the spatial step size by $h$. The equation (\ref{ds}) is accompanied with a
discrete equation for the intermediate layer as:\newline
\begin{gather}
c\left( {\left( \theta ^{n}\right) _{i}^{j+1}-\left( \theta ^{n}\right)
_{i}^{j}}\right) /\tau +c_{n}\left( {\left( \theta ^{n}\right) _{i+1}^{j+%
\frac{1}{2}}-\left( \theta ^{n}\right) _{i-1}^{j+\frac{1}{2}}}\right) /{2h}%
+\sum\limits_{k,m}g_{mkn}\left( \theta ^{m}\right) _{i}^{j+\frac{1}{2}%
}\left( {\left( \theta ^{k}\right) _{i+1}^{j+\frac{1}{2}}-\left( \theta
^{k}\right) _{i-1}^{j+\frac{1}{2}}}\right) /{2h}  \label{il} \\
+e_{n}\left( {\left( \theta ^{n}\right) _{i+2}^{j+\frac{1}{2}}-2\left(
\theta ^{n}\right) _{i+1}^{j+\frac{1}{2}}+2\left( \theta ^{n}\right)
_{i-1}^{j+\frac{1}{2}}-\left( \theta ^{n}\right) _{i-2}^{j+\frac{1}{2}}}%
\right) /{\ 2h^{3}}=0  \notag
\end{gather}
\qquad \qquad

To support the results of further simulations to be presented here we would
recover some mathematical ingredients of our work. Let $(u^{n})_{i}^{j}$ $%
=u^{n}(x_{i},t^{j})$ denotes the grid function obtained from the
exact solution $u^{n}(x,t)$ \ of the Cauchy problem for the cKdV
system calculated at the grid points $x_{i}$, $t^{j}$. Then the
discrete solution $(\theta ^{n})_{i}^{j}$ of the system
(\ref{ds}), (\ref{il} ) defines the discrepancy
vector $V^{j}$ (residual function) with the components $(v^{n})_{i}^{j}=(%
\theta ^{n})_{i}^{j}-(u^{n})_{i}^{j}.$ Introduce the $L_{2}$ norm for $V^{j}$%
:
\begin{equation}
V^{j+1}=\left( \sum\limits_{i}\sum\limits_{n}\left[ \left( v^{n}\right)
_{i}^{j}\right] ^{2}\,h\,\,\,\right) ^{\frac{1}{2}}
\end{equation}

The convergence of the scheme (\ref{ds}), (\ref{il} ) at $(\tau
,h\rightarrow 0)$ follows from the

\begin{theorem}
Let the solution $u^{n}(x,t)$ of the Cauchy problem for the cKdV system with
$t\in \lbrack 0,\infty );x\in (-\infty ,\infty )$ exists with the third
continuous derivative with respect to time $t$, and the fifth derivative
with respect to $x$. Then the following inequality takes place.
\begin{equation} \label{in}
\left\| V^{j+1}\right\| \leq \frac{1-\sqrt{1-\frac{4M\left|
g_{m,k}^{n}\right| _{max}}{h^{\frac{3}{2}}}\,\,M\,O\/\left( \tau
\,+h^{2}\right) }}{\frac{2M\,\left| g_{m,k}^{n}\right| _{max}}{h^{\frac{3}{2}%
}}}\,\cong \,M\,O\,\left( \tau ^{2}\,+h^{2}\right) ,
\end{equation}
where $M$ is a constant independent of $\tau $, $h$ and $\left|
g_{m,k}^{n}\right| _{max}=\max\limits_{1\leq n,m,k\leq L}(\left|
g_{m,k}^{n}\right| )$; $L$ is quantity of modes in (\ref{coupledkdv})$.$
\newline
\end{theorem}

Therefore, the convergence follows from the inequality (\ref{in}).
The proof of the theorem is placed in the Appendix. We give the
proof for more simple case then the only one of the numeric scheme
equations (\ref{ds}) is used in simulations. The proof for the
scheme (\ref{ds})-(\ref{il}) with a half-step is analogous but
significantly more cumbersome, and we decided to give not it here,
taking in mind publishing in the forthcoming special mathematical
paper . This scheme and proof may be of interest even in the
scalar (single KdV) case in spite of existence of numerous
versions of such results. We believe the ours is effective as it
follows from the tests and experiments presented here.

\begin{remark}
We have to note that evidently the existence of the fifth
derivative by $x$ and the third derivative by $\tau $ are
unnecessarily rigid conditions. These conditions are used for
simplicity of consideration only, and may be easily weaken. One
can replace them by the condition of the third order derivative by
$x$ and the first order derivative by $x$ existence, then only the
convergence is more slow. Moreover, one can accept a concept of
weak solutions to be applied in applications. Usually such a
concept allow to weaken the requirement of derivative existence on
one order of derivatives; however the author have not verified it
for the problem under consideration because necessity of such a
delicate analysis is not evident.
\end{remark}

\section{Computation and results}

When the coefficients in the Sturm-Liouville equation
(\ref{sturmliouv}) are constant it has very simple general
solution

\begin{center}
$Z^{n}=B_{n} Sin\left(\frac{n\pi z}{h}\right),\,\,\,n=1,2,3,...,L $ \,\,\,\,%
\\[0pt]
\end{center}
which tends to zero at boundaries. The eigenvalues
\begin{equation}  \label{c-coef}
c_{n}=\frac{Nh}{n\pi},n=1,2,3,...L.
\end{equation}
have the sense of linear IGW velocities. Normalization is determined by $%
\int\limits_{0}^{h}\left(Z^{n}\right)^{2}N^{2}dz=1 \,\,$ and
gives\newline

\begin{center}
$B_{n}=\left(\frac{2}{N^{2}h}\right)^{.5}$, Hence
\end{center}

\begin{equation}
Z^{n}(z )=\left(\frac{2}{N^{2}h}\right)^{.5} Sin\left(\frac{n\pi
z}{h}\right)
\end{equation}
As it was discussed in the Sec.2 one can select the initial
perturbation for the current function (1) which have the general
form at the point $t = 0$

\begin{center}
$\psi(z,x,0) =\sum\limits_{n=1}^{L} Z^{n} (z)
\theta^{n}(x,0)=\varphi(x,z)=\varphi_{1}(x) \varphi_{2}(z).$\\[0pt]
\end{center}

Taking smooth, exponentially localized initial condition for\,\ $%
\varphi_{1}(x)$ while $\varphi_{2}(z)$\,\ that model the impact of
some rotating paddle motion as
\begin{equation}
\,\,\,\,\,\varphi_{1}(x)=\frac{a}{Cosh(\frac{x}{l})}\,\,\,\
and\,\,\,\,\,\,\varphi_{2}(z)=\left(\frac{2}{N^{2}h}\right)^{.5}
Sech(b(z-z_{0}))Tanh(b(z-z_{0}))
\end{equation}
The choice of the form and the constants a, b, l qualitatively
reflects the paddle movement (we restrict the movement by some
isolated pulse) and the numerical value for the amplitude is
estimated also from the description of the experiment of McEwan
(1983a). Then

\begin{equation}  \label{init}
(Z^{j},\psi)=\sum\limits_{n=1}^{L}(Z^{j},Z^{n})
\theta^{n}(x,0)=(Z^{j},\varphi_{2}(z))\,\,\,\varphi_{1}(x).
\end{equation}

\begin{equation}  \label{scal}
(Z^{j},Z^{n})=\int\limits_{0}^{h}N^{2}\,\,Z^{j}\,\,Z^{n}dz=\{1,\,\,
(j=n)\},\{0,\,\,(j\neq n)\}
\end{equation}
\begin{equation}  \label{coef}
(Z^{j},\varphi_{2}(z))=\int\limits_{0}^{h}N^{2}\,Z^{j}\,\varphi_{2}(z)dz.
\end{equation}
For $h=0.25 m \,\,and \,\, N=1.23 s^{-1}$, taking the first five
nonzero modes with coefficients from (\ref{coef}) and the numbers
(2,4,6,8,10) while the odd numbers give zero, together with (\ref{scal}) and putting in (%
\ref{init}) to obtain $\theta(x,0)$. The estimation of the modes
energy satisfactory keep that of the initial perturbation. Then
using (1) we get the initial perturbation $\psi(z,x,0)$. We give
cross section plot at the middle of the tank Fig.(2) to show the
antisymmetry and the percentage of energy content in the initial
perturbation model \newline

\begin{center}
\epsfig{file=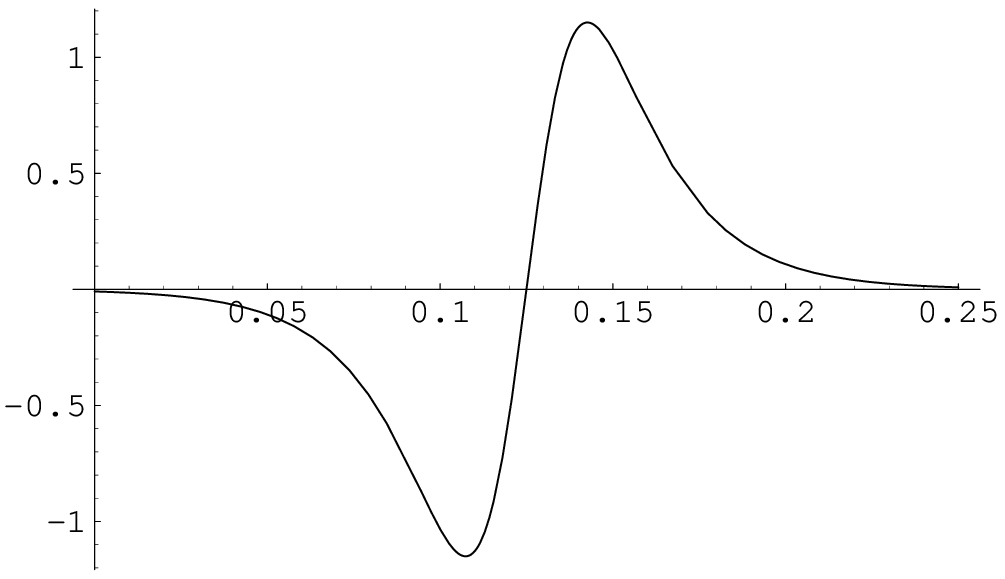, height=4 cm, width=8cm ,clip=,angle=0}
\epsfig{file=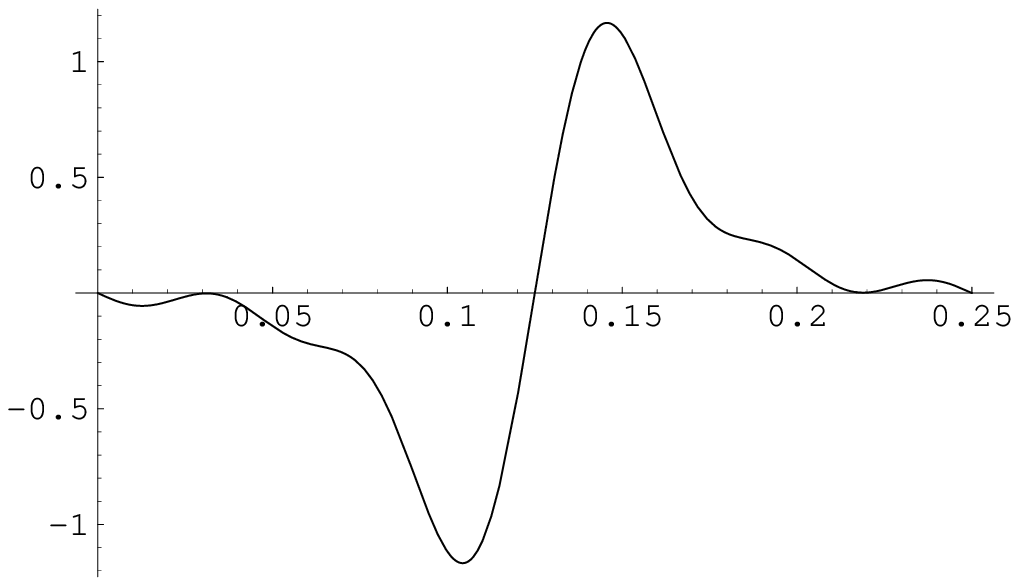, height=4 cm, width=8cm
,clip=,angle=0}\\[0pt]
{{\hspace*{10 mm} Fig.2 $\varphi_{2}(z)$ initially and after
Fourier representation } }
\end{center}
For the wave profile after time t we solve numerically system
(\ref {ckdv1}) with coefficients from (\ref{coef1}),
(\ref{c-coef}) to obtain $\theta^{n}(x,t)$ then using (1) for
$\psi(z,x,t)$ evaluation. For the chosen constants the calculated
coefficients of nonlinear, dispersion
terms and modes propagation velocities are picked together in the table\\[0pt]
 \begin{center}
\begin{tabular}{cccccccccccccccccccccccccc}
\multicolumn{5}{c}{$g^{2}_{mk}$} &
\multicolumn{5}{c}{$g^{4}_{mk}$} &   \\
m$\backslash$k & 2 & 4 & 6 & 8 & 10 &  & m$\backslash$k & 2 & 4 &
6 & 8 & 10
&   \\
2 & 0 & 72.3 & 0 & 0 & 0 &  & 2 & 28.9 & 0 & 57.8 & 0 & 0  \\
4 & 28.9 & 0 & 202.4 & 0 & 0 &  & 4 & 0 & 0 & 0 & 144.5 & 0  \\
6 & 0 & 130 & 0 & 390.1 & 0 &  & 6 & 0 & 0 & 0 & 0 & 260   \\
8 & 0 & 0 & 289 & 0 & 635.7 &  & 8 & 0 & 57.8 & 0 & 0 & 0   \\
10 & 0 & 0 & 0 & 505.7 & 0 &  & 10 & 0 & 0 & 144.5 & 0 & 0   \\
\end{tabular}
\end{center}
\begin{center}
\begin{tabular}{cccccccccccccccccccccccccc}
\multicolumn{5}{c}{$g^{6}_{mk}$} &
\multicolumn{5}{c}{$g^{8}_{mk}$} &     \\
m$\backslash$k & 2 & 4 & 6 & 8 & 10 &  & m$\backslash$k & 2 & 4 &
6 & 8 & 10
  \\
2 & 0 & 33.7 & 0 & 53 & 0 &  & 2 & 0 & 0 & 36.1 & 0 & 50.6  \\
4 & 48.2 & 0 & 0 & 0 & 125.2 &  & 4 & 0 & 57.8 & 0 & 0 & 0  \\
6 & 0 & 0 & 0 & 0 & 0 &  & 6 & 65 & 0 & 0 & 0 & 0  \\
8 & -19.3 & 0 & 0 & 0 & 0 &  & 8 & 0 & 0 & 0 & 0 & 0    \\
10 & 0 & 24 & 0 & 0 & 0 &  & 10 & -36 & 0 & 0 & 0 & 0
\end{tabular}
\end{center}
\begin{center}
\begin{tabular}{ccccc cc ccccc cc ccccc cc ccccc}
\multicolumn{5}{c}{$g^{10}_{mk}$} &   \\
m$\backslash$k & 2 & 4 & 6 & 8 & 10  \\
2 & 0 & 0 & 0 & 37.6 & 0 \\
4 & 0 & 0 & 63.6&0&0  \\
6 & 0 & 78&0&0&0 \\
8 & 80.9&0&0&0&0  \\
10 & 0 & 0 & 0 & 0 & 0
\end{tabular}
\end{center}
\begin{center}
$d_{22}$ =$.00004$,\,\,\,$d_{44}$ =$.000005$,\,\,\,$d_{66}$ =$.000002$,\,\,\,%
$d_{88}$=$.000001$,\,\,\,\,\,\,\,\,$d_{1010}$=$.0000003$,\\[0pt]
$c_{2}$=$.05$,\,\,\,$c_{4}$=$.025$ ,\,\,\,$c_{6}$=$.016$,\,\,\,$c_{8}$=$.012$%
,\,\,\,$c_{10}$=$.0098$\\[0pt]
\end{center}

For the case of linearized equations the wave profile after time
t=0.02 s may be obtained directly from the same system if one put
nonlinear constants to zero. The evaluation result is shown in the
following contour plot Fig.(3) for the upper half of the tank.
\begin{center}
\epsfig{file=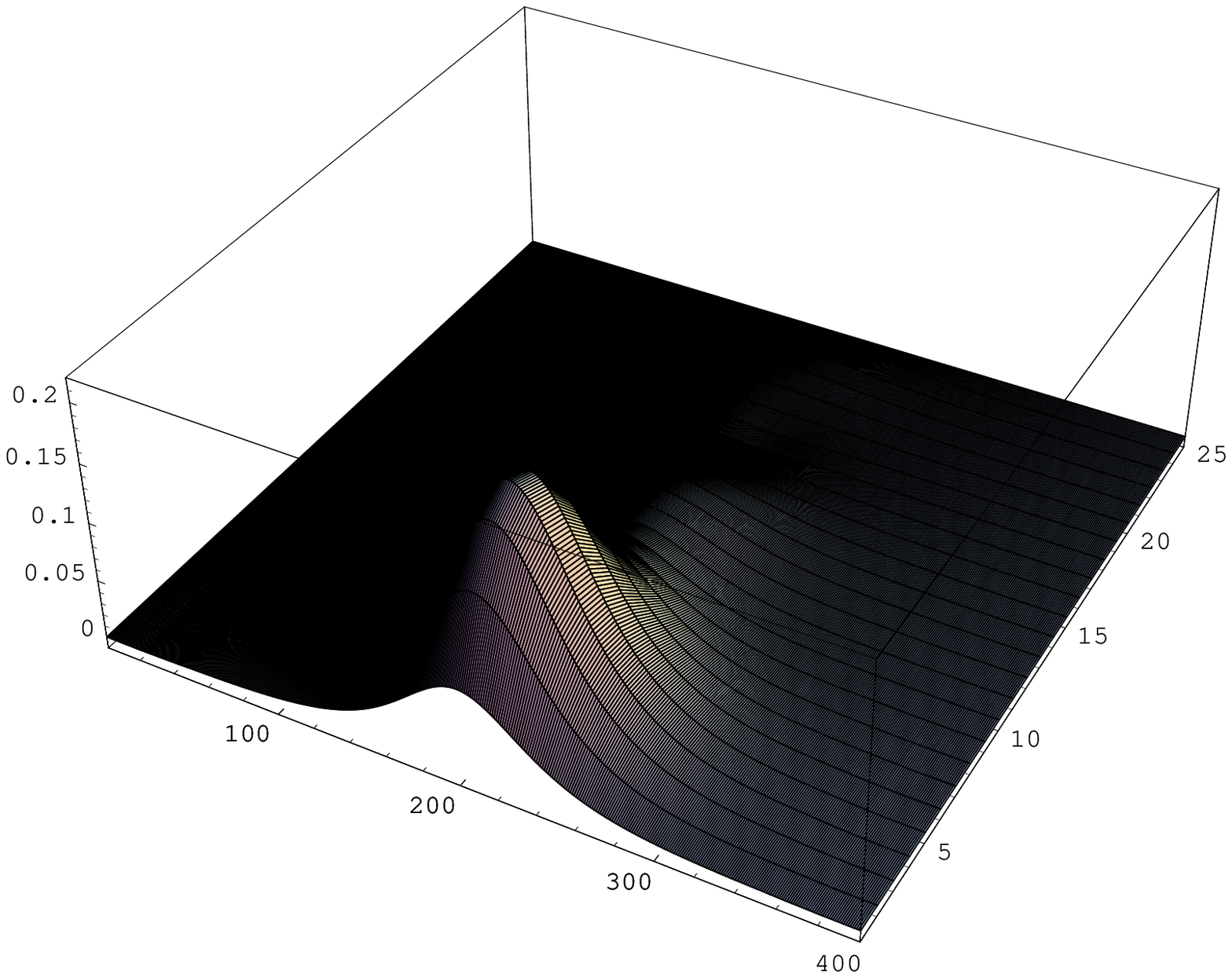, height=8cm, width=10cm ,clip=,angle=0}\\[0pt]
{\hspace*{10 mm} Fig.3 The wave profile 3-dimensional plot in
linear case with dispersion for the upper half of the tank (12.5
cm in z and 50 cm in x directions) }
\end{center}
For the nonlinear case we show one-dimensional (x) profiles
Fig.(4.a,b,c,d,e) for the second, forth, sixth, eighth and tenth
modes respectively
\begin{center}
\epsfig{file=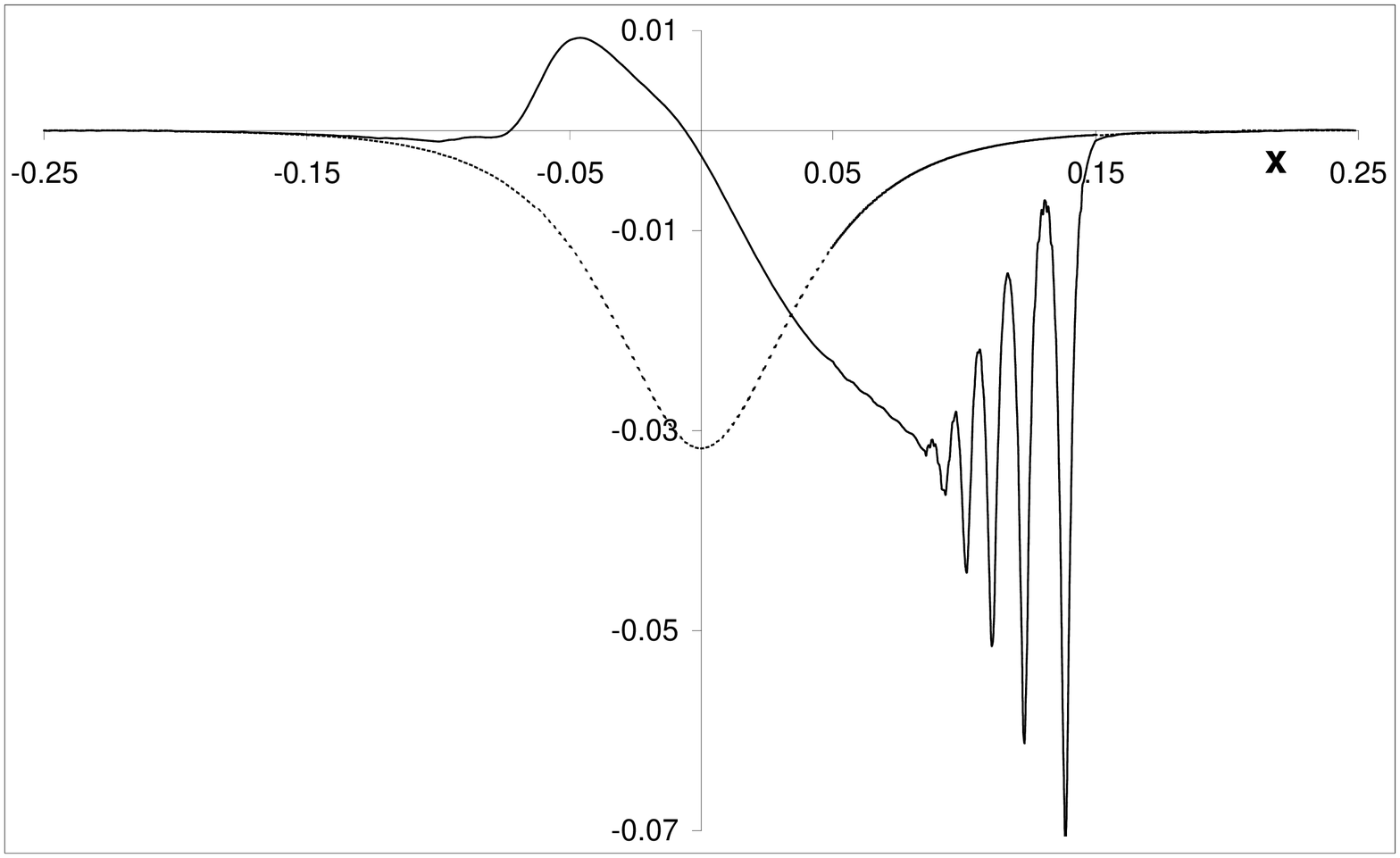, height=3.9 cm, width=14cm ,clip=,angle=0}\\[0pt]
{\hspace*{3 mm} Fig.(4.a) Second mode, initially (dotted) and at
t=0.02
sec(continuous). }\\[0pt]

\epsfig{file=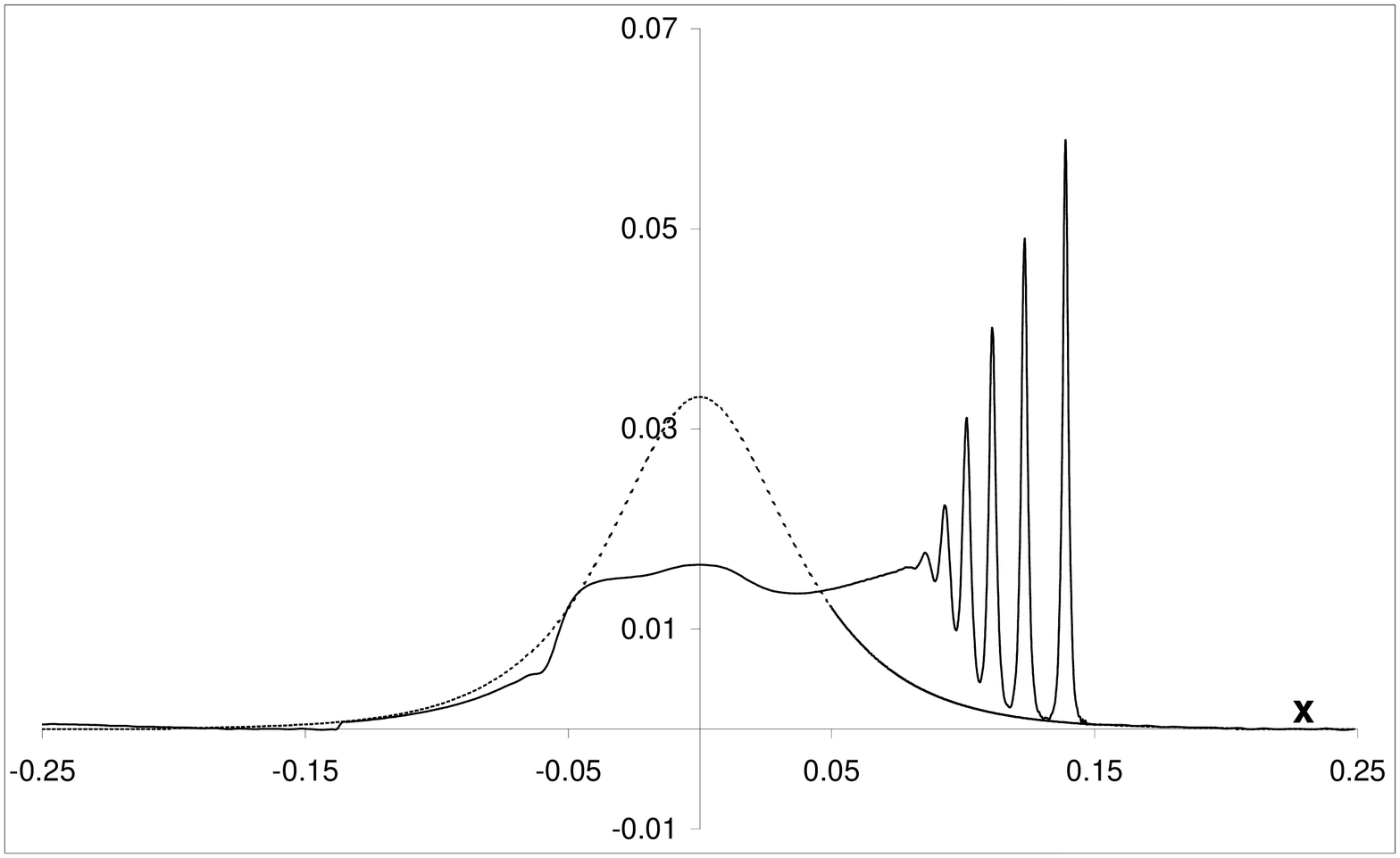, height=3.9 cm, width=14cm ,clip=,angle=0}\\[0pt]
{\hspace*{3 mm} Fig.(4.b) Fourth mode, initially (dotted) and at
t=0.02
sec(continuous). }\\[0pt]

\epsfig{file=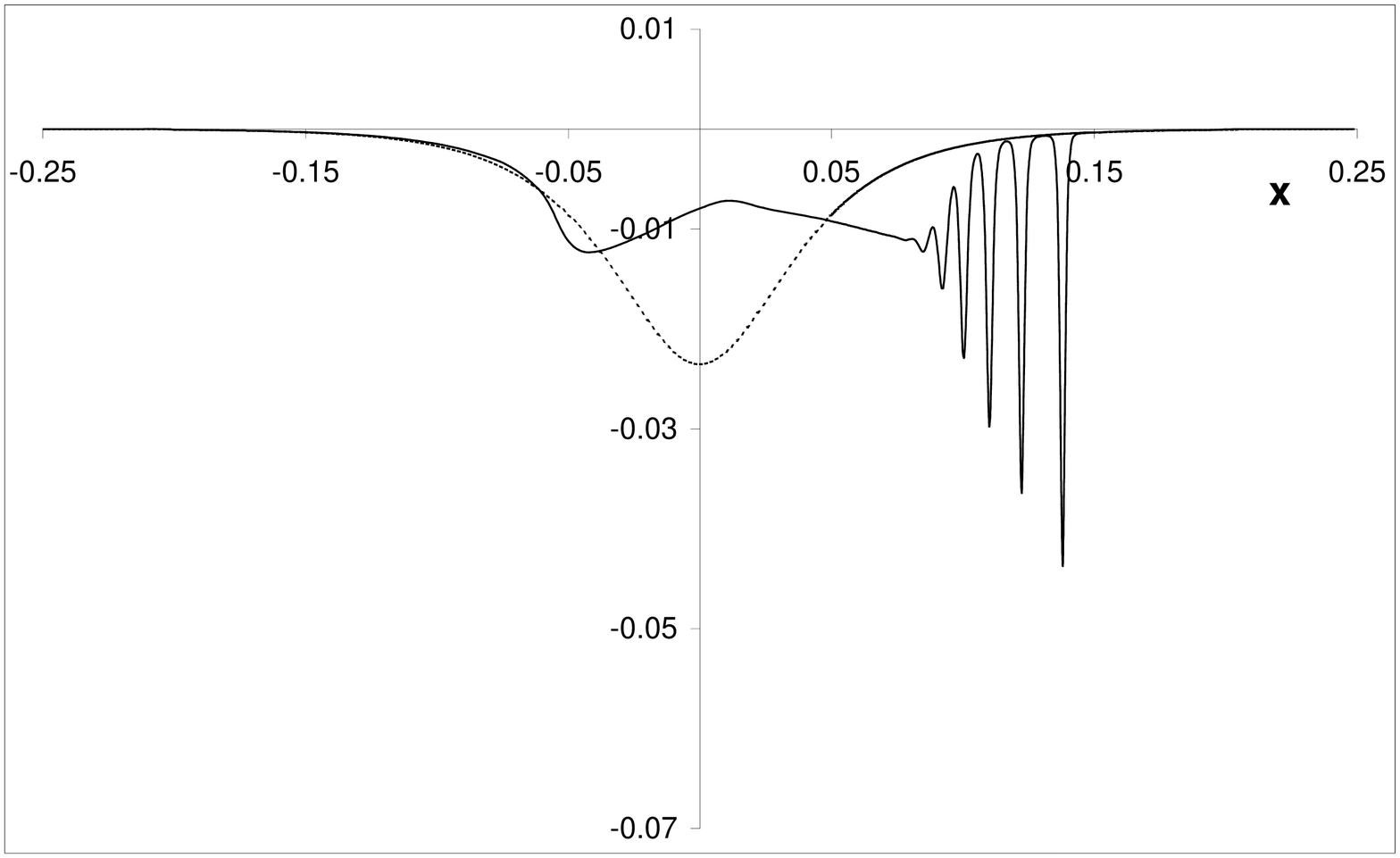, height=3.9 cm, width=14cm ,clip=,angle=0}\\[0pt]
{\hspace*{3 mm} Fig.(4.c) Sixth mode, initially (dotted) and at
t=0.02
sec(continuous). }\\[0pt]

\epsfig{file=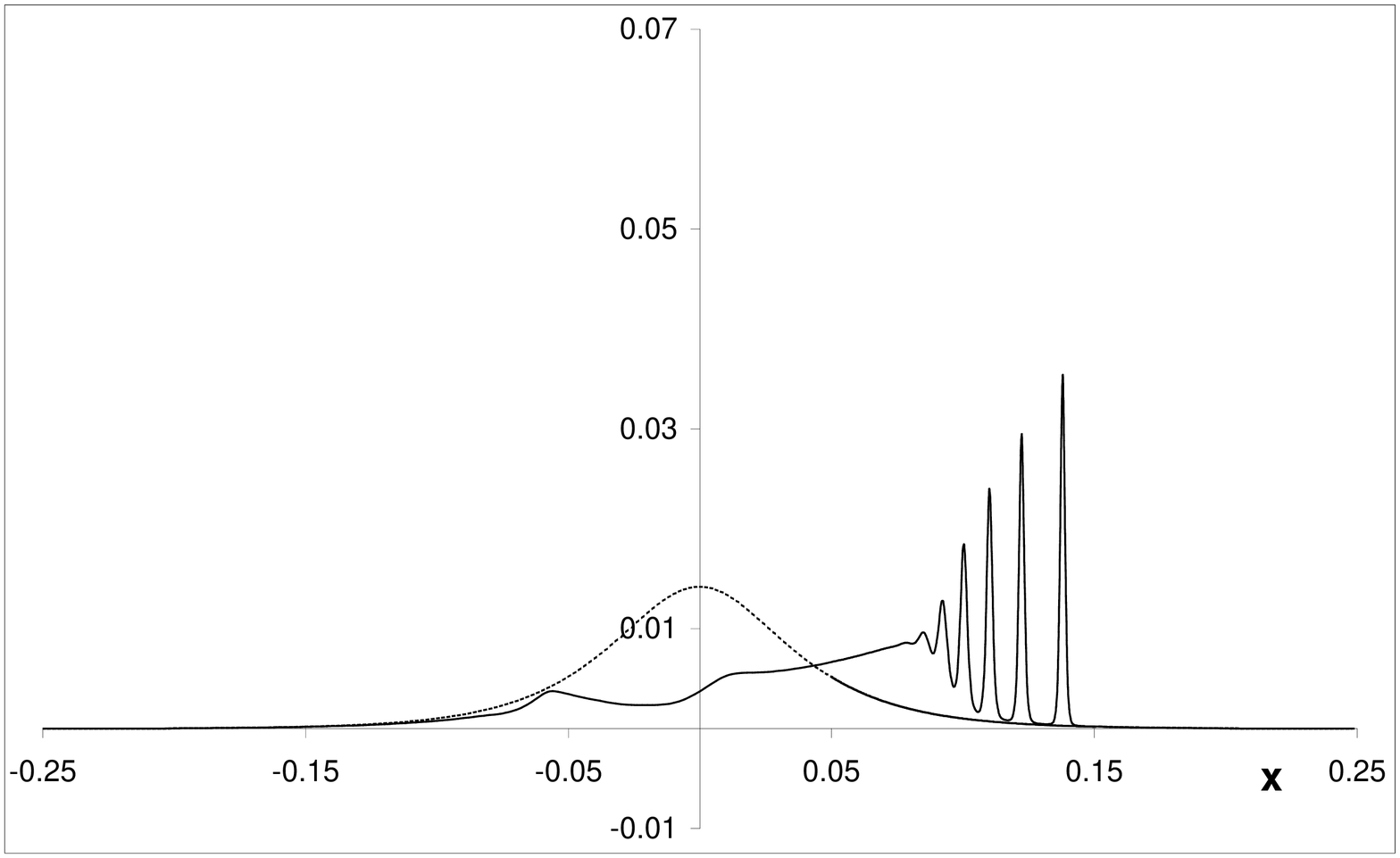, height=3.9 cm, width=14cm ,clip=,angle=0}\\[0pt]
{\hspace*{3 mm} Fig.(4.d) Eighth mode, initially (dotted) and at
t=0.02
sec(continuous). }\\[0pt]

\epsfig{file=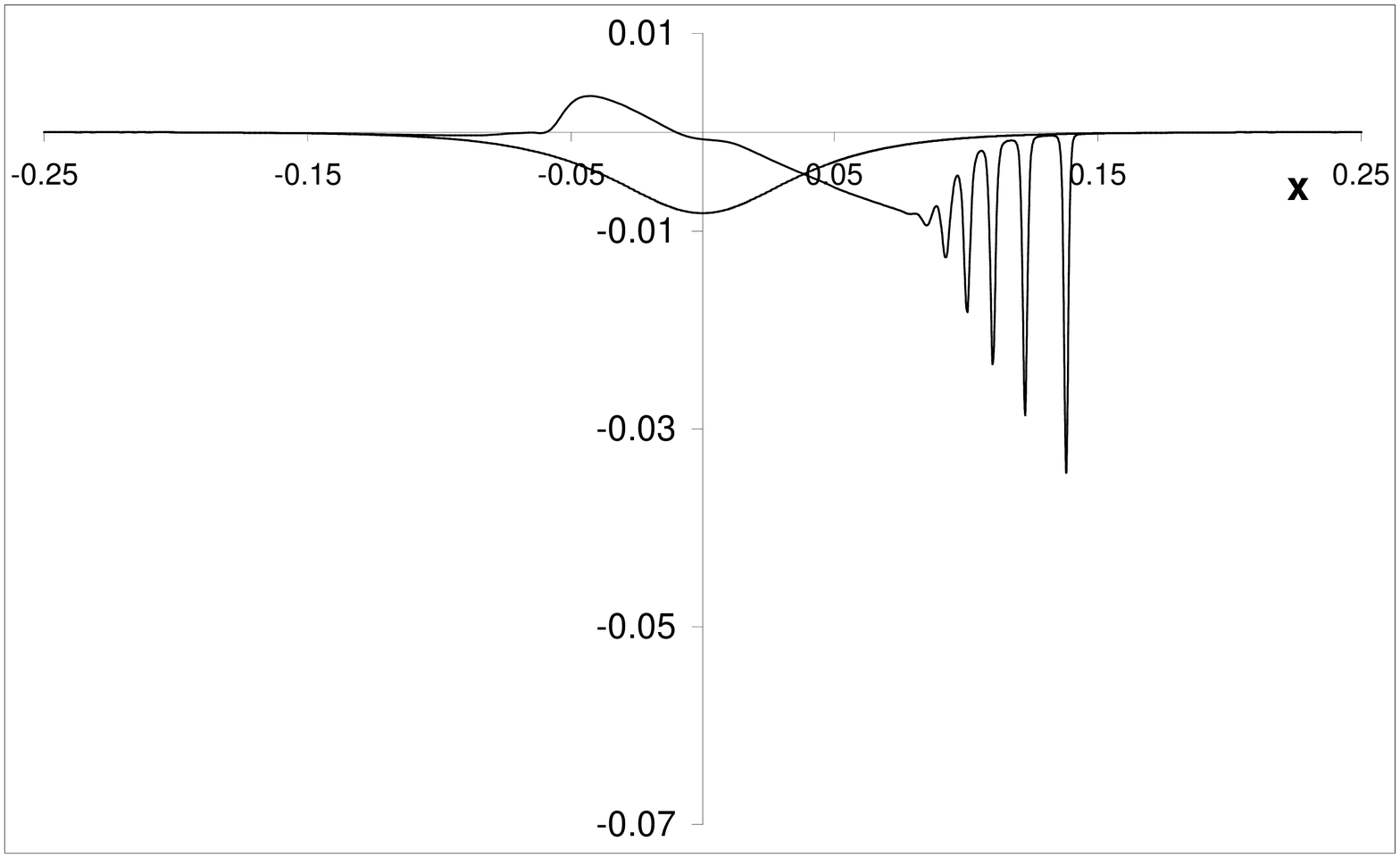, height=3.9 cm, width=14cm ,clip=,angle=0}\\[0pt]
{\hspace*{3 mm} Fig.(4.e) Tenth mode, initially (dotted) and at
t=0.02
sec(continuous). }\\[0pt]

{\hspace*{5 mm} Fig.4 One-dimensional (x) plots of the wave
profile in nonlinear case.}
\end{center}
The contour plot for the wave profile at $t=0.02$ sec is presented
below (Fig.5) for the same half of the tank.
\begin{center}
\epsfig{file=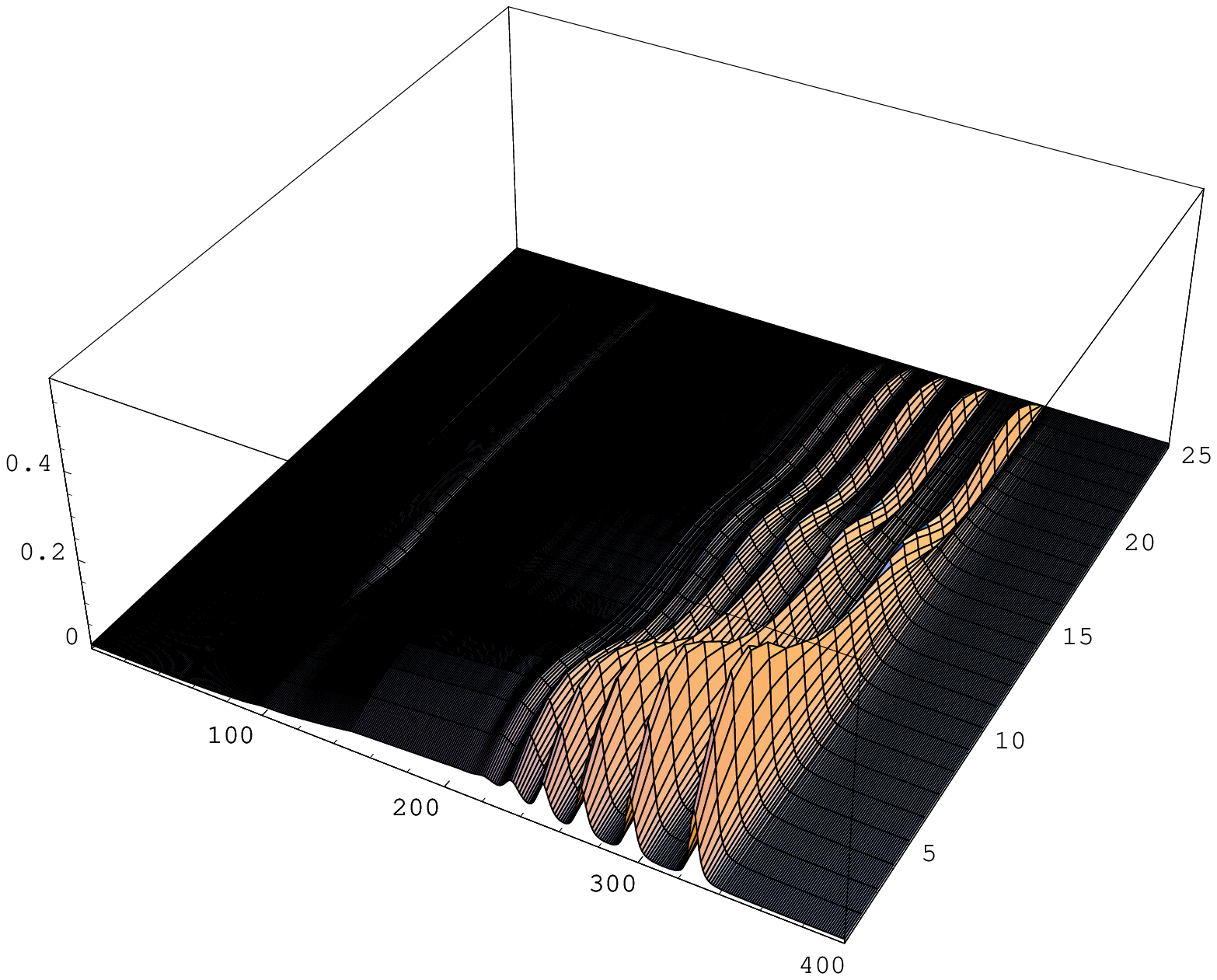, height=14cm, width=18  cm ,clip=,angle=0}\\[0pt]
{\ Fig.4 Three dimensional (x,y,z) plot of the wave profile for
the upper
half of the tank.\\[0pt]
The horizontal numbering indicates the number of points used in
plot while
the vertical\\[0pt]
show the actual z profile of the waves }
\end{center}
All the
plots for the modes and for the sum (current function) show the
behavior that is typical for the multisolitonic perturbation. It
looks likely a decay of the initial condition to solitons in the
single KdV equation theory. The difference between linear
dispersion and nonlinear looks crucial. Even within the short time
interval we present here the horizontal scale of the disturbance
radically changes. The final scale is defined by constants of the
cKdV system and the amplitude of initial condition. The process of
the wave propagation is accompanied by interaction that imply the
energy transfer between modes. This phenomenon may be considered
as a possible reason for the vertical fine structure generation
(Ref. Leble (1991) ) The combination of the fine structures may
explain the experiments of McEwan (1983 a,b).

\section{Conclusion}

We conclude with the following remarks about restrictive
assumptions we made and perspectives. First in the framework of
clarity of the plotting and interpretation we considered only one
direction of propagating waves. We would repeat that one can
elongate the time of simulations to have more interactions but
such picture looks cumbersome.

The relatively small number of modes is taken into account because
it conserves most the energy. The number may be increased but at
the times we consider the effect of small-vertical-scales modes
generation may be neglected. Returning to the general significance
in the theory of the Euler equations validity mentioned in the
introduction, one can easily put the important question. The
question arises when we analyze the expressions for the constants
in the of equations cKdV system. In a limit of the homogeneous
medium, or $N\rightarrow 0$, the nonlinear constant increase while
the dispersion and velocities go to zero. It means the growth of
solitons number and more strong interaction between modes. The
discussion of this phenomenon and ideas of adequate regularization
will be considered in the forthcoming publication.

\section{Appendix A \newline
Theoretical description and mathematical problem.}
Consider Euler
equations that describe IGW in two dimensions (x,z) stratified
water  \smallskip
\begin{equation} \label{Eul2}
\begin{array} {cc}
u_{x}+w_{z}=0 \\
\rho _{o}\,u_{t}=-\rho _{o}\left( \underline{v},\underline{\nabla
}\right) u-p_{x} \\
\rho _{o}\,w_{t}= \rho _{o}\left( \underline{v},\underline{\nabla
} \right) w-p_{z}-\rho ^{^{\prime }}\,g \\
T_{t}^{^{\prime }}+w \overline{T}\,_{z}\,=-\left( \underline{v},%
\underline{\nabla }\right) T^{^{\prime }}
\end{array}
\end{equation}
where u, v are velocity components, $\rho _{o}\,$\ is the density,
$p$ is the pressure, $\rho ^{^{\prime }}\,g$\thinspace\ is the
body force due to stratifications, $\overline{T}\,_{z}$ is the
vertical background temperature gradient and $T^{^{\prime }}$ is
the temperature variables.

Combining equations in (\ref{Eul2}) and using the state equation
for liquid $\rho ^{^{\prime
}}=-\rho _{o}\alpha T^{^{\prime }},$ $\alpha $ $=-\frac{\rho _{z}}{\rho \,%
\overline{T}\,_{z}}$is the coefficients of thermal expansion $\,\,\overline{T%
}\,_{z}=\frac{N^{2}}{\alpha g}\,$we obtain
\begin{equation}\label{vl-vel}
 \Delta w_{tt}+N^{2}w_{xx}-N^{2}\left[ \left( \underline{v},\underline{%
\nabla }\right) \int_{0}^{t}w\,dt\right] _{xx}+\left[ \left( \underline{v},%
\underline{\nabla }\right) \int w_{z}\,dx\right] _{txz}+\left[
\left( \underline{v},\underline{\nabla }\right) w\right] _{txx}=0
\end{equation}
Going to the dimensionless variables by the following formulas

$x_{i}=\lambda _{i}\,x_{i}^{\prime },\,t=\frac{2\,\pi }{\overline{N}\,\beta }%
\,t^{\prime },\,u=\frac{\lambda _{z}\,\overline{N\,}\,}{2\pi }u^{\prime },w=%
\frac{\beta \,\lambda _{z}\,\overline{N\,}\,\,}{2\pi }w^{\prime
},\,\alpha \,T^{\prime }=T,\,N=\frac{\overline{N}\,}{2\pi
}N^{\prime }$

So $\,\left( \partial w/\partial x_{i}\right) =\left( 1/\lambda
_{i}\right) \left( \partial w/\partial x_{i}^{\prime }\right) \ ,\
\ \ \ \left( \partial w/\partial t\right) =\left(
\overline{N\,\,}\,\beta /2\pi \right) \ \left(
\partial w/\partial t^{^{\prime }}\right) \ \ \ \ \,$

Substitute in (\ref{vl-vel}) and omit primes for simplicity

$\lambda _{z}\left( \frac{\bar{N}\beta }{2\pi }\right) ^{3}\left( \frac{%
w_{xx}}{\lambda _{x}^{2}}+\frac{w_{zz}}{\lambda _{z}^{2}}\right)
_{tt}+\left( \frac{\bar{N}}{2\pi }\right) ^{3}\left( \beta
N\right)
^{2}\left( \frac{w_{xx}}{\lambda _{x}}\right) -\left( \frac{\bar{N}}{2\pi }%
\right) ^{3}\frac{\left( \beta N\right) ^{2}}{\lambda _{x}}\left[
u\int\nolimits_{0}^{t}w_{x}dt+w\int\nolimits_{0}^{t}w_{z}dt\right]
_{xx}$

$+\left( \frac{\bar{N}\beta }{2\pi }\right) ^{3}\left\{ \frac{\lambda _{z}}{%
\lambda _{x}^{2}}\left[ uw_{x}+ww_{z}\right] _{xxt}+\frac{1}{\lambda _{z}}%
\left[ uw_{z}+w\left( \int w_{z}\,dx\right) _{z}\right]
_{xzt}\right\} =0$

Introducing the current function , $w\prime =-\sigma \psi
_{x},\,u\prime =\sigma \psi _{z}$ ,\thinspace Integrate w.r.t. x
\begin{equation} \label{psi}
 \psi _{zztt}+N^{2}\psi _{xx}=-\beta ^{2}\psi _{xxtt}-\sigma \left[ \psi
_{z}\psi _{xz}-\psi _{x}\psi _{zz}\right] _{zt}+  \sigma
N^{2}\left[ \psi _{z}\int\nolimits_{0}^{t}\psi _{xx}dt-\psi
_{x}\int\nolimits_{0}^{t}\psi _{xz}dt\right] _{x}
\end{equation}

Substitute in (\ref{psi}) by $\psi =\underset{m}{\sum }\,z^{m}\theta ^{m}\,,\,\,$%
multiply\thinspace\ by$\,\ \,z^{n}$\thinspace , \thinspace
inegrate\thinspace \thinspace \thinspace w.r.t.\thinspace\ z and
using The
separation of variables that gives
\begin{equation} \label{separ}
\ z_{zz}^{n}=-\frac{N^{2}}{c_{n}^{2}} \,z^{n}\medskip
\end{equation}
So we have
\begin{equation} \label{theta}
\theta _{tt}^{n}-c_{n}^{2}\,\theta
_{xx}^{n}=\frac{c_{n}^{2}\,\beta ^{2}}{ N^{2}}\,\theta
_{xxtt}^{n}+\sigma \,c_{n}^{2}\underset{m,k}{\sum }
a_{m,k}^{n}\left( \theta ^{m}\theta _{x}^{k}\right) _{t}+\left[
b_{m,k}^{n}\,\theta ^{m}\theta _{t}^{k}+e_{m,k}^{n}\,\theta _{x}^{m}\overset{%
t}{\underset{0}{\int }}\theta _{x}^{k}dt\right] _{x}
\end{equation}

where$\,\ a_{m,k}^{n}=N^{2}\overset{h}{\underset{-h}{\int }}\left( \frac{-1}{%
c_{k}^{2}}+\frac{1}{c_{m}^{2}}\right) z_{z}^{m}z^{k}z^{n}dz
,b_{m,k}^{n}=\frac{-N^{2}}{c_{k}^{2}
}\overset{h}{\underset{-h}{\int }}z_{z}^{m}z^{k}z^{n}dz,\,e_{m,k}^{n}=N^{2}\overset{h}{\underset{-h}{\int }}%
z^{m}z_{z}^{k}z^{n}dz\,\ \ \ \ \ \ \ \ \ \ \ \ \ \ \ $

The equations of the separated propagated modes are obtained as follow\\
Let \ $\theta _{t}^{n}=u^{n}\,,\,\ \,c_{n}\theta _{x}^{n}=v^{n}$

So (\ref{theta}) becomes
\begin{equation} \label{u-v}
\begin{array}{cc}
 u_{t}^{n}-c_{n}u_{x}^{n}=\frac{c_{n}^{3}\beta ^{2}}{N^{2}}
v_{xxx}+Nonlinear\,terms \\
v_{t}^{n}-c_{n}u_{x}^{n}=0
 \end{array}
\end{equation}

Using the projection operator P, $P_{+}\,=\frac{1}{2}\left(
\begin{array}{c}
1\,\ \ \ \ \,\ \ \ 1+\frac{c^{2}\beta ^{2}}{2N^{2}}\partial _{x}^{2} \\
1-\frac{c^{2}\beta ^{2}}{2N^{2}}\partial _{x}^{2}\,\ \ \ \ \ \ \ \
1
\end{array}
\right) , P_{-}=\frac{1}{2}\left(
\begin{array}{c}
1\,\ \ \ \ \,\ \ \ -1-\frac{c^{2}\beta ^{2}}{2N^{2}}\partial _{x}^{2} \\
-1+\frac{c^{2}\beta ^{2}}{2N^{2}}\partial _{x}^{2}\,\ \ \ \ \ \ \
\ 1
\end{array}
\right) $

\begin{equation} \label{u-v-phi}
\begin{array}{cc}
P_{+}\left(
\begin{array}{c}
u^{n} \\
v
\end{array}
\right) =\ \left(
\begin{array}{c}
\varphi ^{n+} \\
k\varphi ^{n+}
\end{array}
\right) ,\,P_{-}\left(
\begin{array}{c}
u \\
v
\end{array}
\right) =\left(
\begin{array}{c}
\varphi ^{-} \\
-k\varphi ^{-}
\end{array}
\right) \\
u=\varphi ^{n+}+\varphi ^{n-}\,,\,\ \ v=\left( \varphi
^{n+}-\varphi ^{n-}\,\right) -\frac{c^{2}\beta
^{2}}{2N^{2}}\partial _{x}^{2}\left( \varphi ^{n+}-\varphi
^{n-}\right)
\end{array}
\end{equation}

Operating  $P_{+}\,,\,P_{-}\,$on (\ref{u-v}) and using
(\ref{u-v-phi}) $\ $we obtain the equations for the separated
modes $\varphi ^{n+},\,\varphi ^{n-}$
\begin{equation} \label{phi+}
\begin{array}{cc}
\varphi _{t}^{n+}-c_{n}\varphi _{x}^{n+}-\frac{c^{3}\beta
^{2}}{2N^{2}} \varphi _{xxx}^{n+}-\frac{\sigma
\,c_{n}^{2}}{2}\underset{m,k}{\sum \,} a_{m,k}^{n}\left( \int
\left( \varphi ^{+}+\varphi ^{-}\right) ^{m}dt.\frac{
\left( \varphi ^{+}-\varphi ^{-}\right) ^{k}}{c_{k}}\right) _{t}   \\
+\left[ b_{m,k}^{n}\int \left( \varphi ^{+}+\varphi ^{-}\right)
\,^{m}dt.\left( \varphi ^{+}+\varphi ^{-}\right)
^{k}+e_{m,k}^{n}\frac{ \left( \varphi ^{+}-\varphi ^{-}\right)
^{m}}{c_{m}}\,\overset{t}{\underset{0 }{\int }}\frac{\left(
\varphi ^{+}-\varphi ^{-}\right) ^{k}}{c_{k}}dt\right]_{x} \ \
\end{array}
\end{equation}
\bigskip\
\begin{equation} \label{phi-}
\begin{array}{cc}
\varphi _{t}^{n-}+c_{n}\varphi _{x}^{n-}+\frac{c^{3}\beta ^{2}}{2N^{2}}%
\varphi _{xxx}^{n-}-\frac{\sigma \,c_{n}^{2}}{2}\underset{m,k}{\sum \,}%
a_{m,k}^{n}\left( \int \left( \varphi ^{n+}+\varphi ^{n-}\right) ^{m}dt.%
\frac{\left( \varphi ^{+}-\varphi ^{-}\right) ^{k}}{c_{k}}\right) _{t} \\
+\left[ b_{m,k}^{n}\int \left( \varphi ^{+}+\varphi ^{-}\right)
\,^{m}dt.\left( \varphi ^{+}+\varphi ^{-}\right) ^{k}+e_{m,k}^{n}\frac{%
\left( \varphi ^{n+}-\varphi ^{n-}\right) ^{m}}{c_{m}}\,\overset{t}{%
\underset{0}{\int \,}}\frac{\left( \varphi ^{+}-\varphi ^{-}\right) ^{k}}{%
c_{k}}dt\right] _{x}
\end{array}
\end{equation}
\bigskip

Let $\varphi ^{n+}=\Psi _{t}^{n+},\,\ \varphi ^{n-}=\Psi
_{t}^{n-}\,\ \ \ \ \ \ $ So (\ref{phi+}), (\ref{phi-})  becomes
\begin{equation} \label{psi-}
\begin{array}{cc}
\Psi _{t}^{n+}-c_{n}\Psi _{x}^{n+}=\frac{c^{3}\beta
^{2}}{2N^{2}}\Psi
_{xxx}^{n+}+\frac{\sigma \,c_{n}^{2}}{2}\underset{m,k}{\sum \,\frac{a_{m,k}^{n}}{c_{k}}}%
\left( \Psi ^{+}+\Psi ^{-}\right) ^{m}\left( \Psi ^{+}-\Psi
^{-}\right)
_{x}^{k}  \\
+ b_{m,k}^{n}\left( \Psi ^{+}+\Psi ^{-}\right) \,^{m}.\left(
\varphi ^{+}+\varphi ^{-}\right)
_{x}^{k}+\frac{e_{m,k}^{n}}{c_{k}c_{m}}\left( \Psi ^{+}-\Psi
^{-}\right) _{x}^{m}\,\left( \Psi ^{+}-\Psi ^{-}\right) ^{k}
\end{array}
\end{equation}

\bigskip
\begin{equation} \label{psi-}
\begin{array}{cc}
\Psi _{t}^{n+}+c_{n}\Psi _{x}^{n+}=-\frac{c^{3}\beta
^{2}}{2N^{2}}\Psi
_{xxx}^{n+}+\frac{\sigma \,c_{n}^{2}}{2}\underset{m,k}{\sum \,\frac{a_{m,k}^{n}}{c_{k}}}%
\left( \Psi ^{+}+\Psi ^{-}\right) ^{m}\left( \Psi ^{+}-\Psi
^{-}\right)
_{x}^{k}  \\
+ b_{m,k}^{n}\left( \Psi ^{+}+\Psi ^{-}\right) \,^{m}.\left(
\varphi ^{+}+\varphi ^{-}\right)
_{x}^{k}+\frac{e_{m,k}^{n}}{c_{k}c_{m}}\left( \Psi ^{+}-\Psi
^{-}\right) _{x}^{m}\,\left( \Psi ^{+}-\Psi ^{-}\right) ^{k}
\end{array}
\end{equation}

In next work we will consider both directed modes. In this work
  we evaluate only for short time to describe the phenomena. so we
 consider only one direction that have the form
\begin{equation} \label{ckdv2}
\Psi _{t}^{n}+c_{n}\Psi _{x}^{n}+\sigma \underset{m,k}{\sum
\,g_{m,k}^{n}}\Psi ^{m}\Psi _{x}^{k}+\beta ^{2}d_{n}\,\Psi
_{xxx}^{n}=0,
\end{equation}
  $\sigma ,\,\beta ^{2}\,$are scale parameters ,
\begin{equation} \label{coef2}
a_{m,k}^{n}=\frac{\sigma
N^{2}c_{n}^{2}}{2}\overset{h}{\underset{-h}{\int }}\left[ \left(
\frac{-1}{c_{m}^{2}}+\frac{2}{c_{k}^{2}}\right) Z^{k}Z_{z}^{m}-
\frac{1}{c_{m}c_{k}} Z^{m}Z_{z}^{k}\right]
z^{n}dz,\,d_{n}=\frac{c_{n}^{3}}{2N^{2}}
\end{equation}
\section{Appendix B \newline
Proof of the convergence.}

Now we realize that the finite-difference method used gives us a
correct solution of the problem if the time and space steps is
enough small and we show how we can easily estimate an error of
the numerical method. At first step we prove  that a solution
$(\theta ^n _i )^j$ of  finite-difference equations (\ref{ds}),
 \ref{il}) converges to some solution of equation (\ref{coupledkdv}) as $\tau , h \to 0$.

We give  here a  proof for more simple case for brevity when the
scheme is based on  equation (\ref{ds}). The proof for equations
(\ref{ds}), \ref{il}) is analogous but more cumbersome. Therefore,
now we consider the difference equations
\begin{gather}
\frac{\left( \theta ^{n}\right) _{i}^{j+1}-\left( \theta ^{n}\right) _{i}^{j}%
}{\tau }+c_{n}\frac{\left( \theta ^{n}\right) _{i+1}^{j}-\left( \theta
^{n}\right) _{i-1}^{j}}{2h}+\sum\limits_{m,k}g_{m,k}^{n}\left( \theta
^{m}\right) _{i}^{j}\frac{\left( \theta ^{k}\right) _{i+1}^{j}-\left( \theta
^{k}\right) _{i-1}^{j}}{2h}  \label{finite-difference} \\
+e_{n}\frac{\left( \theta ^{n}\right) _{i+2}^{j}-2\left( \theta ^{n}\right)
_{i+1}^{j}+2\left( \theta ^{n}\right) _{i-1}^{j}-\left( \theta ^{n}\right)
_{i-2}^{j}}{2h^{3}}=0  \notag
\end{gather}
 Here $h$ is a space grid step, $i$ is a discrete space variable, $j$ is  a discrete
time, $\tau $ is a time step, $n$ is a mode number. We suppose
that an exact solution is a smooth one and therefore
\begin{eqnarray}
 \frac {(\theta ^n_i )^{j+1}-(\theta ^n_i)^j}{\tau}= \theta
^n_t |_{x=ih} +O(\tau) \qquad
 \frac {(\theta ^n_{i+1 })^{j}-(\theta ^n_{i-1})^j}{2 h}= \theta
^n_x |_{x=ih} +O(h^2)        \label{approximation} \\
 \frac {(\theta ^n_{i+2 })^{j}-2 (\theta ^n_{i+1 })^{j}+2 (\theta
^n_{i-1 })^{j}-(\theta ^n_{i-2})^j}{2 h^3}= \theta ^n_{xxx} \notag
|_{x=ih} +O(h^2)
\end{eqnarray}
In this way, equations (\ref{finite-difference}) approximates
equations (\ref{coupledkdv}) with an error of the order of
$O(\tau+h^2)$.

We know that a single KdV equation $ u_{t}+uu_{x}+u_{xxx}=0 $ has
the conservation law $\int_{-\infty}^{\infty} u^{2}(x,t)dx = const
$. So we can consider a solution of a single KdV equation as an
element of $L_2$-space, parameterized by  time $t$. Equations
(\ref{coupledkdv}) are of the same structure as a simple KdV
equation, so it is naturally to develop the proof in the
functional space with $L_2$-norm.

Let $\left( \theta ^{n}\right) ^{j}$ denote a column
 $$
 \left(
\theta ^{n}\right) ^{j}\equiv \left(
\begin{array}{c}
\cdots \\
\left( \theta ^{n}\right) _{i-2}^{j} \\
\left( \theta ^{n}\right) _{i-1}^{j} \\
\left( \theta ^{n}\right) _{i}^{j} \\
\left( \theta ^{n}\right) _{i+1}^{j} \\
\left( \theta ^{n}\right) _{i+2}^{j} \\
\cdots
\end{array}
\right), \qquad n=1,2,\cdots,l
$$
and let  $\Theta ^j$ denote a column consisting of columns $\left(
\theta ^{n}\right) ^{j}$ introduced above:
$$\Theta ^{j}\equiv \left(
\begin{array}{c}
\left( \theta ^{1}\right) ^{j} \\
\left( \theta ^{2}\right) ^{j} \\
\left( \theta ^{3}\right) ^{j} \\
\cdots           \\
\left( \theta ^{l}\right) ^{j}
\end{array}
\right)
$$

Let $u^n(x,t)$ be an exact solution of (\ref{coupledkdv}). Then
$(u_i^n)^j \equiv u^n(x_i=ih,t=j \tau)$ is an exact solution in
the point $x_i$ of the grid at the moment $j \tau$.  Let
 $$
 \left(
u ^{n}\right) ^{j}\equiv \left(
\begin{array}{c}
\cdots \\
\left( u ^{n}\right) _{i-2}^{j} \\
\left(u ^{n}\right) _{i-1}^{j} \\
\left( u  ^{n}\right) _{i}^{j} \\
\left( u  ^{n}\right) _{i+1}^{j} \\
\left( u  ^{n}\right) _{i+2}^{j} \\
\cdots
\end{array}
\right), \qquad n=1,2,\cdots,l,  \qquad \qquad \text{and} \quad U
^{j}\equiv \left(
\begin{array}{c}
\left( u ^{1}\right) ^{j} \\
\left( u ^{2}\right) ^{j} \\
\left( u ^{3}\right) ^{j} \\
\cdots \\
\left( u ^{l}\right) ^{j}
\end{array}
\right)
$$
The difference $V^j=\Theta^j-U^j $ is an error of the numerical
method. The $L_2$-norm of $V^j$ is
\begin{equation}
\left\| V ^{j}\right\| =\left(
\sum\limits_{n=1}^l\sum\limits_{i=-\infty}^{+\infty}\left[ \left(
(v ^{n})_i^j\right) ^{2}h\right] \right) ^{\frac{1}{2}}.
\label{L2norm}
\end{equation}

The difference solution $\left( \theta ^{n}\right) _{i}^{j}$
converges to the  exact solution $\left( u^{n}\right) _{i}^{j}$ if
$\left\| V^{j}\right\| \rightarrow 0$ as $\tau ,h\rightarrow 0$.

We substitute (\ref{App-Error}) into (\ref{finite-difference}) and
obtain an equation for the error $(v^n)_i^j$
\begin{gather}
\frac{\left( v^{n}\right) _{i}^{j+1}-\left( v^{n}\right) _{i}^{j}}{\tau }%
+c_{n}\frac{\left( v^{n}\right) _{i+1}^{j}-\left( v^{n}\right) _{i-1}^{j}}{2h%
}+\sum\limits_{m,k}\left( g_{m,k}^{n}\left( u^{m}\right) _{i}^{j}\frac{%
\left( v^{k}\right) _{i+1}^{j}-\left( v^{k}\right) _{i-1}^{j}}{2h}\right.
+g_{m,k}^{n}\left( v^{m}\right) _{i}^{j}\frac{\left( u^{k}\right)
_{i+1}^{j}-\left( u^{k}\right) _{i-1}^{j}}{2h}  \label{App-VU} \\
+g_{m,k}^{n}\left( v^{m}\right) _{i}^{j}\,\left. \frac{\left( v^{k}\right)
_{i+1}^{j}-\left( v^{k}\right) _{i-1}^{j}}{2h}\right) \,\,+e_{n}\frac{\left(
v^{n}\right) _{i+2}^{j}-2\left( v^{n}\right) _{i+1}^{j}+2\left( v^{n}\right)
_{i-1}^{j}-\left( v^{n}\right) _{i-2}^{j}}{2h^{3}}=-\left( \frac{\left(
u^{n}\right) _{i}^{j+1}-\left( u^{n}\right) _{i}^{j}}{\tau }\right. +  \notag
\\
\,c_{n}\frac{\left( u^{n}\right) _{i+1}^{j}-\left( u^{n}\right) _{i-1}^{j}}{%
2h}+\sum\limits_{m,k}g_{m,k}^{n}\left( u^{m}\right) _{i}^{j}\frac{\left(
u^{k}\right) _{i+1}^{j}-\left( u^{k}\right) _{i-1}^{j}}{2h}+e_{n}\left.
\frac{\left( u^{n}\right) _{i+2}^{j}-2\left( u^{n}\right) _{i+1}^{j}+2\left(
u^{n}\right) _{i-1}^{j}-\left( u^{n}\right) _{i-2}^{j}}{2h^{3}}\right) \,
\notag
\end{gather}
Pick out a linear part of expression (\ref{App-VU}) and introduce
for convenience an operator $T^{j}$ \ by the expression
\begin{gather}
\left( v^{n}\right) _{i}^{j}-\tau \,\left( c_{n}\frac{\left( v^{n}\right)
_{i+1}^{j}-\left( v^{n}\right) _{i-1}^{j}}{2h}\sum\limits_{m,k}\left(
g_{m,k}^{n}\left( u^{m}\right) _{i}^{j}\frac{\left( v^{k}\right)
_{i+1}^{j}-\left( v^{k}\right) _{i-1}^{j}}{2h}+g_{m,k}^{n}\left(
v^{m}\right) _{i}^{j}\left. \frac{\left( u^{k}\right) _{i+1}^{j}-\left(
u^{k}\right) _{i-1}^{j}}{2h}\right) \right. \right. + \\
\,\,\,\,\,\,\,\,\,\,\,\,\,\,\,\,\,\,\,\,\,\,\,\,\,\,\,\,\,\,\,\,e_{n}\left.
\frac{\left( v^{n}\right) _{i+2}^{j}-2\left( v^{n}\right)
_{i+1}^{j}+2\left( v^{n}\right) _{i-1}^{j}-\left( v^{n}\right)
_{i-2}^{j}}{2h^{3}}\right) =\sum\limits_{r=-\infty}^\infty\left(
T^{j+1}\right) _{ir}\,\left( v^{n}\right)
_{r}^{j},\,\,n=1,2,3,...l.  \notag
\end{gather}

Because  $\left( u^{n}\right) _{i}^{j}$ is an exact solution of
differential equations,   the right part of (\ref {App-VU})  is a
small of the order $O\left( \tau +h^{2}\right).$ Therefore, we can
rewrite (\ref{App-VU}) in the following way
\begin{gather*}
\left( v^{n}\right) _{i}^{j+1}-\sum\limits_{r}\left(
T^{j+1}\right) _{ir}\left( v^{n}\right) _{i}^{j}+\tau
\sum\limits_{m,k}g_{m,k}^{n}\left( v^{m}\right)
_{i}^{j}\frac{\left( v^{k}\right) _{i+1}^{j}-\left(
v^{k}\right) _{i-1}^{j}}{2h}=\tau \left( f^{n}\right) _{i}^{j}, \\
\left( f^{n}\right) _{i}^{j}=-
\sum\limits_{m,k=1}^lg_{m,k}^{n}\left( v^{m}\right)
_{i}^{j}\frac{\left( v^{k}\right) _{i+1}^{j}-\left( v^{k}\right)
_{i-1}^{j}}{2h}+O\left( \tau +h^{2}\right)
\end{gather*}

We need an estimate of a norm of $T^j$. It is valid
\begin{lemma}
\label{lemma1} If $u^{n}(x,t)$ is a function with a bounded
derivative $\left| \frac{\partial }{\partial x}u^{n}(x,t)\right|
$, then for any $b>0$ and for all
\begin{equation}
\tau \leq bh^{6},  \label{condition}
\end{equation}
there  exists a constant $A$, independent of $\tau $, $h$, $j$,
such that $\left\| T^{j}\right\| _{s}\preceq \exp (A\tau )$.
\end{lemma}

Here $\left\| T^{j}\right\| _{s}$ is a spectral norm of the operator $T^{j}$%
, i.e., the norm of $T^{j}$ that is induced by $L_{2}$-norm
$\left\| V^{j}\right\| $. We will omit the subscribe ''$s$'' in
the following and will simply write $ \left\| T^{j}\right\| $. The
proof of  Lemma \ref{lemma1} is a standard one, but cumbersome.
Therefore we do not give the proof of  Lemma \ref{lemma1} here and
recommend  reader  books  \cite{Lancas}, \cite{Marchuk}.

Let $F^j$ be the column constructed from $(f^n_i)^j$ in the same
manner as $\Theta^j $ constructed from $(\theta ^n_i) ^j $.
 We  need a norm of $ F^{j}$;
simple estimation gives
\begin{gather}
\left\| F^{j}\right\| =\left( \sum \limits_{n=1}^l
\sum\limits_{i=-\infty}^{\infty}\left[ \left( f^{n}\right)
_{i}^{j}\right] ^{2}h\right) ^{\frac{1}{2}}= G \,\left(
\sum\limits_{n=1}^l\sum\limits_{i=-\infty}^\infty\left[
\sum\limits_{m,k=1}^l\,\left( v^{m}\right)
_{i}^{j}\frac{\left( v^{k}\right) _{i+1}^{j}-\left( v^{k}\right) _{i-1}^{j}}{%
2h}\right] ^{2}h\right) ^{\frac{1}{2}} +O(\tau+h^2) \leq  \label{Norm-f} \\
G  \sum\limits_{n=1}^l \left(
\sum\limits_{m=1}^l\sum\limits_{i=-\infty}^\infty\left[ \left(
v^{m}\right) _{i}^{j}\right]
^{2}h\sum\limits_{k=1}^l\sum\limits_{i=-\infty}^\infty\left[
\left( v^{k}\right) _{i}^{j}\right] ^{2}h\ \right) ^{\frac{1}{2}}\frac{1}{h^{%
\frac{3}{2}}}+O\left( \tau +h^{2}\right)\leq \frac{Gl}{h^{\frac{3}{2}}}%
\left( \left\| V^{j}\right\| \right) ^{2}+O\left( \tau
+h^{2}\right)     \notag
\end{gather}
where   $G =\max\limits_{1\leq n,m,k\leq l}\left|
g_{m,k}^{n}\right| $.

So, using the operator $T^{j}$, we can rewrite (\ref{App-VU}) as
\begin{equation*}
V^{j+1}=T^{j+1}V^{j}+\tau F^{j}.
\end{equation*}
The following  estimate is valid
\begin{equation}
\left\| V^{j+1}\right\| \leq \left\| T^{j+1}\right\| \left\|
V^{j}\right\| +\tau \left\| F^{j}\right\| \label{Norm-v}
\end{equation}
If we consequently substitute (\ref{Norm-v}) into itself, we get
%\begin{multline}
\begin{eqnarray}
\left\| V^{j+1}\right\| \leq \left\| T^{j+1}\right\| \left\|
V^{j}\right\| +\tau \left\| F^{j}\right\| \leq \left\|
T^{j+1}\right\| \left\| T^{j}\right\| \left\| V^{j-1}\right\|
+\tau (\left\| T^{j+1}\right\| \left\|
F^{j-1}\right\| +\left\| F^{j}\right\| )\leq  \label{longformular} \\
\left\| T^{j+1}\right\| \left\| T^{j}\right\| \left\|
T^{j-1}\right\| \left\| V^{j-2}\right\| +\tau (\left\|
T^{j+1}\right\| \left\| T^{j}\right\| \left\| F^{j-2}\right\|
+\left\| T^{j+1}\right\| \left\| F^{j-1}\right\|
+\left\| F^{j}\right\| \leq  \notag \\
e^{A\tau j}\left\| V^{0}\right\| +\tau (e^{A\tau (j-1)}\left\|
F^{0}\right\| +e^{A\tau (j-2)}\left\| F^{1}\right\| +...\left\|
F^{j}\right\| )\leq  \notag
\\
e^{A\tau j}\left\| V^{0}\right\| +M\max_{k\leq j}(\left\|
F^{k}\right\| ) \leq \notag \\
e^{A\tau j}\left\| V^{0}\right\| +M\left\| F^{j+1}\right\| ,
\notag \\ \qquad M=\tau \frac{e^{At_{\max }}-1}{e^{A\tau }-1}.
\notag
\end{eqnarray}
%\end{multline}
We have used a known formula for the sum of a geometric series.
Here  $t_{\max }$ is the time of simulation: $0\leq t\leq t_{\max
}$ .

Substitution of (\ref{Norm-f}) instead of $\left\| F^{j}\right\| $
into  (\ref{longformular}) yields the inequality
\begin{equation*}
\left\| V^{j+1}\right\| \leq e^{A \tau j}\left\| V^{0}\right\|
+M\left( \frac{ Gl }{h^{\frac{3}{2}}}\left( \left\|
V^{j+1}\right\| \right) ^{2}\right) +O\left( \tau +h^{2}\right)
\end{equation*}

Because of the initial conditions we know  exactly, we have
$(v^n_i)^0=0$. So, taking into consideration $\ \left\|
V^{0}\right\| =0$ we obtain  the solution of the inequality
\begin{equation*}
\left\| V^{j+1}\right\| \leq \frac{1-\sqrt{1-\frac{4M Gl }{h^{\frac{3}{2}}}MO\/\left( \tau +h^{2}\right) }}{%
\frac{2M Gl }{h^{\frac{3}{2}}}}\cong MO\left( \tau +h^{2}\right)
\end{equation*}
i.e. $\left\| V^{j+1}\right\| \rightarrow 0$ as $\tau $,
$h\rightarrow 0$. The convergence is proved.

The error of the method considered is of order $O\left( \tau +h^{2}\right) $%
, if the solution under consideration is enough smooth one. The
method converges for any $\tau, \, h \to 0$, but the rate of
converging is dependent of relation between $\tau$ and $h$: the
more $\frac {\tau}{h^6}$ the better. If the solution is not
smooth, then we must consider the solution in the sense of a
distribution theory; this case is not study yet.

The proof of the convergence for the scheme (\ref{ds}), (\ref{il}
) is analogous, but more cumbersome and tedious. The error of the
method (\ref{ds}), (\ref {il} ) is $O\left( \tau ^{2}+h^{2}\right)
$, and instead of condition (\ref{condition} )  the condition
$\tau \leq b  h^{4}$  is used. The main idea of the proof is the
same one: at small $\tau$, $h$ the error due to difference
approximation of the nonlinear terms influence on the result more
less then the errors due to difference approximation of dispersion
terms; and all difficult estimates only are necessary to show this
fact. Therefore,  choice of the optimal step $tau $ may be
accomplished with the help of consideration of simple difference
equation with dispersion term only. We had used this fact when
made numerical experiments.

\end{document}